\documentclass[prb,twocolumn]{revtex4-1}

\usepackage[english]{babel}
\usepackage{graphicx}
\usepackage{epstopdf}
\usepackage{amssymb}
\usepackage{amsmath}
\usepackage{mathrsfs}

\begin{document}

\title{Odd Frequency Density Waves}

\author{Yaron Kedem}
\affiliation{Center for Quantum Materials, Nordic Institute for Theoretical Physics (NORDITA), Roslagstullsbacken 23, S-106 91 Stockholm, Sweden}

\author{ Alexander V. Balatsky}
\affiliation{Institute for Materials Science, Los Alamos National Laboratory, Los Alamos, NM 87545, USA }
\affiliation{Center for Quantum Materials, Nordic Institute for Theoretical Physics (NORDITA), Roslagstullsbacken 23, S-106 91 Stockholm, Sweden}

\begin{abstract}

A new type of hidden order in many body systems is explored. This order appears in states which are analogues to charge density waves, or spin density waves, but involve anomalous particle-hole correlations that are odd in relative time and frequency. These states are shown to be inherently different from the usual states of density waves. We discuss two methods to experimentally observe the new type of pairing where a clear distinction between odd and even correlations can be detected: (i) by measuring the density-density correlation, both in time and space and (ii) via the conductivity which, according to the Kubo formula, is given by the current-current correlation. An order parameter for these states is defined and calculated for a simple model, illuminating the physical nature of this order.

\end{abstract}

\maketitle

\section{Introduction}
The conventional way to consider ordered many body systems is the Landau-Ginzburg formalism of phase transition that is centered on order parameter as the attribute of order. That appearing order is classified as the anomalous correlations that describe the emergent state. For example, in charge density wave the order parameter is
\begin{equation}\label{kOP}
\Delta_\mathbf{q}(\mathbf{k}) = \langle a^\dagger_{\mathbf{k}+\mathbf{q},\alpha} a_\mathbf{k,\alpha} \rangle.
\end{equation}
where $ a^{(\dagger)}_{ \mathbf{k},\alpha}$ is creating (annihilating) an electron with momentum $\mathbf{k}$ and spin $\alpha$. Similarly, the order parameter for spin density waves is given by
\begin{equation}\label{sOP}
\vec {\Delta}_\mathbf{q}(\mathbf{k}) = \langle a^\dagger_{\mathbf{k}+\mathbf{q},\alpha} a_{\mathbf{k}, \beta} \rangle \vec {\sigma}_{\alpha \beta}.
\end{equation}
where $\vec{\sigma}$ is a vector of Pauli matrices. Summation over spin indices is implied, both in (\ref{kOP}) and (\ref{kOP}) and these anomalous correlations are taken at equal time for the involved operators. Static values as written are often taken as the order parameters of the charge (spin) density waves.\cite{Gruner1,Gruner2}

As our view of correlations evolves, it becomes apparent that materials can exhibit composite meta order that significantly expands the old Ginzburg-Landau paradigm.
Examples of new orders include topological order with no local order parameter \cite{top1,top2,top3,top4}. Another extension of the concept of order is the notion of odd frequency or odd-time correlations, first proposed by Berezinskii. \cite{Ber,BA} Odd-frequency order in superconductors and superfluids has been proposed for numerous realizations and most likely to occur in superconducting heterostructures, in nanoscale devices\cite{EfetovRMP,Tanakareview} and in mulitband superconductors\cite{ABSB}. In addition to superconductivity odd time orders were expanded to spin nematics \cite{Nematicodd} and BEC\cite{bec}. Here we wish to further advance the notion of odd time orders by considering the odd in time density wave correlations. This odd time density waves would describe dynamic order that does not manifest itself in any static density wave. The situation can be viewed as a dynamic order that is hidden from the conventional spectroscopies of charge and spin density waves.

\section{Introducing odd time CDW}

The odd time, or odd frequency, orders are characterized by correlations that vanish when $ \langle a^\dagger a \rangle$ expectaion is taken at equal time. We now focus on charge and spin density waves (CDW,SDW). Since the discussion, in our context, for these two types of states, is highly similar, for the sake of simplicity, we will first treat only CDW, described by eq. (\ref{kOP}). Since spin indices are summed over, they are suppressed in all expressions. In order to treat SDW, described by eq. (\ref{sOP}) one simply has to reintroduce the spin indices and multiply by $\vec{\sigma}$ throughout the derivation.

We can generalize the correlation (\ref{kOP}) to include operators $a_\mathbf{k}(\tau)$ which act in Matsubara time $\tau$. The result is the anomalous green function
\begin{eqnarray} \label{kOPt}
\Delta_\mathbf{q} (\mathbf{k},\tau) &=& \left\langle T_\tau \left[ a^\dagger_{\mathbf{k}+\mathbf{q}}(\tau) a_\mathbf{k} (0) \right] \right\rangle
\nonumber\\
&=& \theta(\tau) \left\langle a^\dagger_{\mathbf{k}+\mathbf{q}}(\tau) a_\mathbf{k} (0) \right\rangle
\nonumber\\ &-& \theta(-\tau) \left\langle a_\mathbf{k} (0) a^\dagger_{\mathbf{k}+\mathbf{q}}(\tau) \right\rangle
\end{eqnarray}
where $ T_\tau$ is the time ordering operator and $\theta(x)$ is the Heaviside step function. Since the system is assumed to be invariant to time translation, $ \Delta_\mathbf{q} (\mathbf{k},\tau) $ is a function of the relative time only.

There can be a few possibilities for the time dependency of $ \Delta_\mathbf{q} (\mathbf{k},\tau)$ at short times, $ \tau \rightarrow 0 $. If there is a zero time order, i.e. a time independent part, then we recover the usual form of CDW with non vanishing order parameter given by (\ref{kOP}). Then the time dependence of the correlator at nonzero times would reflect the retardation effects that occur at a later time. If, on the other hand, the equal time part vanishes, but the first order $ \Delta_\mathbf{q} (\mathbf{k},\tau) \sim \tau$ remains finite, the situation is different. In such a case, all the well-known results regarding CDW might not be valid, but nonetheless there will be non-trivial density-density correlations. We are thus facing the possibility to have a nontrivial correlations that have no equal time signature, i.e. a hidden order.

To capture the dynamic nature of the correlation we consider a scenario is when $ \Delta_\mathbf{q} (\mathbf{k},\tau)$ is odd in time:
\begin{equation} \label{oddtime}
\Delta_\mathbf{q} (\mathbf{k},\tau) = - \Delta_\mathbf{q} (\mathbf{k},-\tau).
\end{equation}
In this case it is clear that $\Delta_\mathbf{q} (\mathbf{k})=\Delta_\mathbf{q} (\mathbf{k},0)= 0$. This implies that if the system is described by quantity obeying a condition of the form (\ref{oddtime}), it is in a state that possess a dynamic kind of order. This state might share many of the characteristics, of the usual CDW, by it is fundamentally different in that there is no static density correlations.

\section{Detection of odd time CDW}
Here we outline experimental observation that would test the odd frequency density order. The usual type of CDW, which is described by (\ref{kOP}) is manifested with a modulation of the density
\begin{equation} \label{ncdw}
n(\mathbf{r}) =\langle \Psi^\dagger(\mathbf{r}) \Psi(\mathbf{r}) \rangle = n_0 + n_\mathbf{q} \cos(\mathbf{q} \mathbf{r} + \phi),
\end{equation}
where $\Psi(\mathbf{r}) = \int d\mathbf{k} e^{i \mathbf{k} \mathbf{r}} a_\mathbf{k}$, $n_0$ is the average density, $n_\mathbf{q}$ is the amplitude of the modulation with wave number $\mathbf{q}$ and $\phi$ is an arbitrary phase.
When writing Eq. (\ref{ncdw}) we ignore any density modulations which are not directly related to CDW, so we explicitly assume a uniform density for $n_\mathbf{q} = 0$. Indeed, the uniform density $n(\mathbf{r}) = n_0$  would be the case when the equal time order parameter vanish, $\Delta_\mathbf{q} (\mathbf{k})=0$, as would occur, for example, in the odd frequency case. So one has to study higher order correlations to reveal this kind of order. This can be done by looking at the density correlation, both in space and time
\begin{align} \label{DC}
\langle T_\tau \hat{n}(\mathbf{r},\tau)& \hat{n}(\mathbf{r}',0) \rangle = n^2 \\
&+ \langle T_\tau \Psi^\dagger(\mathbf{r},\tau) \Psi(\mathbf{r}',0) \rangle \langle T_\tau \Psi^\dagger(\mathbf{r}',0) \Psi(\mathbf{r},\tau) \rangle, \nonumber
\end{align}
where we used Wick's theorem and assumed that $n= \langle \hat{n}(\mathbf{r},\tau) \rangle$ is spatially and temporally independent. This assumption involves two important aspects. First the spatial homogeneity means we do not have the usual CDW state, which is manifested by the modulations (\ref{ncdw}). Second the time $\tau$ in $\langle \hat{n}(\mathbf{r},\tau) \rangle$ is the absolute time, or ``center of mass" time, as opposed to $\tau$ in $ \langle T_\tau \hat{n}(\mathbf{r},\tau) \hat{n}(\mathbf{r}',0) \rangle $ which represent a relative time, or some correlation time. Dependency on the relative time can comes from microscopical dynamics. Here we note that there are discussions on states that depend on the overall time. Such a scenario can imply the system is not in a steady state, or even the formation of a time crystal \cite{time,time2}. We do not address those issues in this work.

\begin{figure}
\centering
\includegraphics[trim=9cm 5cm 5cm 3.8cm, clip=true, width=0.48\textwidth]{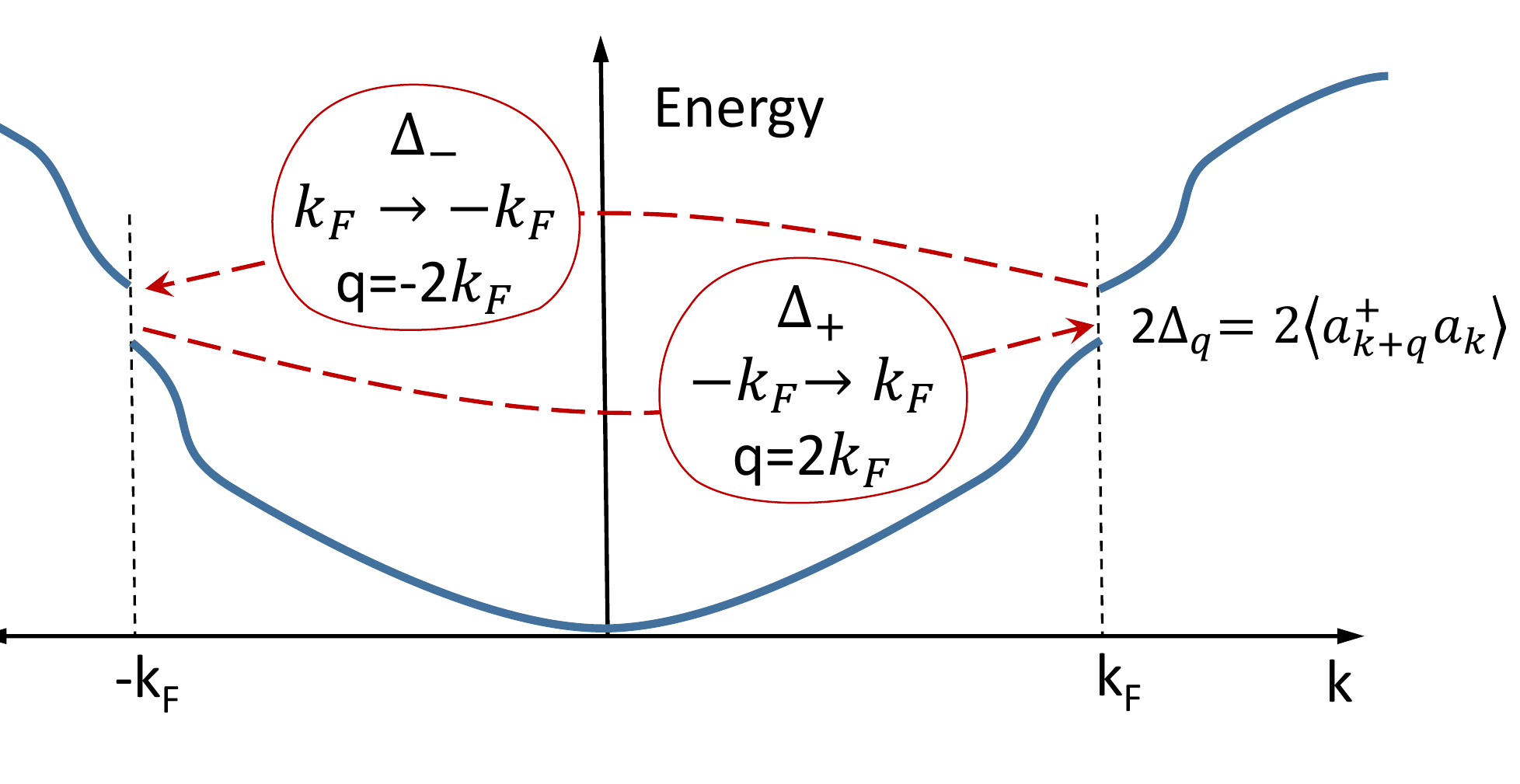}
\caption{ (Color online) An illustration of the typical situation of CDW. Correlations between electrons on opposite sides of the Fermi surface appear and a gap is opened, around the Fermi energy, proportional to the correlation magnitude. Since the involved electrons should be in the vicinity of the Fermi wave vector $\pm k_F$, the momentum difference, which also determine the wave vector of the density modulations, have to be $|\mathbf{q}|=2k_F$. The correlation operator can be written either as annihilation of electron at $k_F$ and creation at $-k_F$ so $\Delta_\mathbf{q}(\mathbf{k}) = \langle a^\dagger_{-k_F } a_{k_F} \rangle$, which make $\mathbf{q}=-2k_F$, or vice versa $\Delta_\mathbf{q}(\mathbf{k}) = \langle a^\dagger_{k_F } a_{-k_F} \rangle$, which makes $\mathbf{q}=2k_F$. Note that for odd time CDW $\Delta_\mathbf{q}(\mathbf{k}) = \Delta_\mathbf{q}(\mathbf{k},\tau =0) =0$ and a gap might not open at $k_F$, but it is still reasonable to assume that correlations with the same momentum properties appear. This assumption is represented in eq (\ref{delta}).
}
\label{band}
\end{figure}

The quantity of interest for us is the second term on the right hand side of (\ref{DC}), $\chi(\mathbf{r},\mathbf{r}',\tau) = \langle T_\tau \hat{n}(\mathbf{r},\tau) \hat{n}(\mathbf{r}',0) \rangle - n^2$. By going over to momentum space we can write it using (\ref{kOPt}):
\begin{equation} \label{fur}
\chi (\mathbf{r},\mathbf{r}',\tau) = \int d \mathcal{K} e^{-i(\mathbf{\tilde{r}}\mathbf{\tilde{k}} + \mathbf{r} \mathbf{q} + \mathbf{r}'\mathbf{q}')} \Delta_\mathbf{q}(\mathbf{k},\tau) \Delta_{\mathbf{q}'}(\mathbf{k}',-\tau),
\end{equation}
where $\mathbf{\tilde{r}} = \mathbf{r} - \mathbf{r}'$, $\mathbf{\tilde{k}} = \mathbf{k} - \mathbf{k}'$ and $\int d \mathcal{K} = d\mathbf{k} d\mathbf{k}' d\mathbf{q} d\mathbf{q}'$ means integration on all momentum variables.
Let us now consider a case similar to the typical scenario for CDW, shown in Fig \ref{band}, where the abnormal correlations occur on the two opposite sides of a Fermi surface. This implies $\mathbf{q} = - 2 \mathbf{k} = \pm 2 \mathbf{k}_f$, where $\mathbf{k}_f$ is the Fermi wave vector. So the order parameter is given by
\begin{align} \label{delta}
\Delta_\mathbf{q}(\mathbf{k},\tau) &= \Delta_+(\tau)\delta(\mathbf{k} - \mathbf{k}_f) \delta(\mathbf{q} + 2 \mathbf{k}_f) \nonumber \\
& + \Delta_-(\tau) \delta(\mathbf{k} + \mathbf{k}_f) \delta(\mathbf{q} - 2 \mathbf{k}_f).
\end{align}
where $\Delta_+(\tau)$ and $\Delta_-(\tau)$ are in principle independent quantities. Here, for simplicity we choose $\delta$-function as the form of $\Delta_\mathbf{q}(\mathbf{k},\tau)$. A more realistic assumption would be that it is sharply peaked function at those values, so that $\Delta_{+ (-)}(\tau) = \int d\mathbf{k} d\mathbf{q} \Delta_\mathbf{q}(\mathbf{k},\tau) $ with integration intervals are around the points $ \mathbf{k}=(-) \mathbf{k}_f,  -q = (-) 2 \mathbf{k}_f$.

Inserting (\ref{delta}) in (\ref{fur}) we have,
\begin{align} \label{xi}
\chi (\mathbf{r},\mathbf{r}',\tau) &= \Delta_+(\tau) \left[ \Delta_+(-\tau)e^{ i 4 \mathbf{R} \mathbf{k}_f }+ \Delta_-(-\tau) \right] \nonumber\\ &+ \Delta_-(\tau) \left[ \Delta_-(-\tau)e^{ -i 4\mathbf{R} \mathbf{k}_f }
+ \Delta_+(-\tau) \right].
\end{align}
The fact that $ \chi (\mathbf{r},\mathbf{r}',\tau) $ depends on the ``center of mass" coordinate $\mathbf{R}=(\mathbf{r}+\mathbf{r}')/2$ is a clear manifestation of the broken translation symmetry associated with CDW. Since we are considering a situation where the average density is uniform, the broken symmetry that can be observed in the density-density correlation is a strong indication to a hidden order residing in the system.

The first two terms in (\ref{fur}) are oscillating with effective wave vector of $4k_f$. Such terms would appear for the usual case of CDW, if one looks at density-density correlations, in addition to terms with a wave vector of $2k_f$, which is also how the density is modulated. The more rapid oscillations of might be difficult to observe in certain systems. Neglecting the oscillating term we are left with a constant, i.e. $\mathbf{r}$ independent term. In the simple case where $\Delta_+(\tau)= \Delta_-(\tau)= \Delta(\tau)$, the sign of this term is given by the time parity of $\Delta(\tau)$, regardless of the sign of $\Delta(\tau)$ itself. Thus, if $\Delta(\tau)= - \Delta(-\tau)$, there will be a negative contribution from $ \chi (\mathbf{r},\mathbf{r}',\tau) $ to the density-density correlation. We can consider a scenario, where this fact can be the basis for experimental evidence for odd time correlation, when the density-density correlation is a measured function of some experimental parameter, like temperature. If the correlation is constant in some region and then starts changing, the direction of the change can tell us whether $ \Delta_\mathbf{q} (\mathbf{k},\tau)$ is odd or even in time, i.e. a decrease in correlation would suggest an odd time CDW. 

Another quantity that can be of interest is the current-current correlations, which according to the Kubo formula\cite{Kubo} can yield the conductivity. Assuming the average current vanishes $ \langle j \rangle = 0$, the current-current correlations are given by (see appendix \ref{japp} for details):
\begin{align} \label{jj1}
\langle T_\tau j_i&(\mathbf{r},\tau) j_j(\mathbf{r}',0) \rangle = \int d \mathcal{K} e^{-i(\mathbf{\tilde{r}}\mathbf{\tilde{k}} + \mathbf{r} \mathbf{q} + \mathbf{r}'\mathbf{q}')} \times \nonumber\\
&{ (\mathbf{k} + \mathbf{k}' + \mathbf{q})_i (\mathbf{k} + \mathbf{k}' + \mathbf{q}')_j \over 4 m^2} \Delta_\mathbf{q}(\mathbf{k},\tau) \Delta_{\mathbf{q}'}(\mathbf{k}',-\tau).
\end{align}
Following the same procedure we used above for $ \chi (\mathbf{r},\mathbf{r}',\tau) $, we consider the typical case, shown in Fig \ref{band}, where $\Delta_\mathbf{q}(\mathbf{k},\tau)$ is given by (\ref{delta}). Inserting (\ref{delta}) into (\ref{jj1}), we get
\begin{align} \label{jj2}
\langle j_i(\mathbf{r},\tau) &j_j(\mathbf{r}',0) \rangle = \nonumber\\
&- { \left(\mathbf{k}_f\right)_i \left(\mathbf{k}_f\right)_j \over m^2} \left[ \Delta_+(\tau) \Delta_-(-\tau) + \Delta_-(\tau) \Delta_+(-\tau)\right].
\end{align}
It is interesting that the rapidly oscillating terms, which appear in eq (\ref{fur}), vanish here because they correspond to zero momentum. The disappearance of a conductivity term, accompanied by the appearance of oscillating density terms seems similar, and might be closely related, to the usual scenario of CDW where the conductivity vanish together with the appearance of density modulation.

\begin{table}
\begin{tabular}{|c c| r r r r || c r|}
\hline
$P$& $T$ & $\Delta_+^+$ & $\Delta_+^-$ & $\Delta_-^+$ & $\Delta_-^-$ & $ \chi (\mathbf{R},\tau) $ & $\langle j j \rangle { m^2 \over k_f^2} $ \\
\hline \hline
+ & + & 1 & 1 &1 &1 & $2+2\cos(4 \mathbf{R} k_f)$ & -2 ~~ \\
+ & - &1 & 1 & -1 &-1 & $-2+2\cos(4 \mathbf{R} k_f)$ & 2 ~~ \\
- & + &1 & -1 & 1 &-1 & $-2-2\cos(4 \mathbf{R} k_f)$ & 2 ~~\\
- & - &1 & -1 & -1 & 1 & $2-2\cos(4 \mathbf{R} k_f)$ & -2 ~~\\
\multicolumn{2}{| r|}{Entang} & 1 & 1 & 1 & -1 & $-2 i \sin(2 \mathbf{R} k_f)$ & 0 ~~\\
\hline
\end{tabular}
\caption{The values of the density-density and the current-current correlation for different parity behaviors of $\Delta_\mathbf{q}(\mathbf{k},\tau)$ according te eq. (\ref{xi}) and (\ref{jj2}). $\Delta_\alpha^\beta$ denotes $\Delta_\alpha( \beta \tau)$, $P$ and $T$ are the reversal operator for time and momentum respectively. The last line refer to a situation where $P$ and $T$ do not have a defined value each, but are interdependent, which is noted as Entang(led). From the observation of these values can one can learn the type, and especially the parity, of electron-hole correlations present in a system.}
\label{par}
\end{table}

In the same way we consider $\Delta_\mathbf{q}(\mathbf{k},\tau)$ to be an odd or even function of time, we can also consider whether it is odd or even in momentum. For the case described by (\ref{delta}), even (odd) parity in momentum implies $\Delta_+(\tau)= (-) \Delta_-(\tau)$. These properties of $\Delta_\mathbf{q}(\mathbf{k},\tau)$ can be formulated by considering $\Delta_\mathbf{q}(\mathbf{k},\tau)$ to be a eigenfunction of two operators: the time reversal $T$, defined as $T \Delta_\mathbf{q}(\mathbf{k},\tau) = \Delta_\mathbf{q}(\mathbf{k},-\tau) $ and the momentum reversal, defined as as $P \Delta_\mathbf{q}(\mathbf{k},\tau) = \Delta_{-\mathbf{q}}(-\mathbf{k},\tau) $. For each operator, an eigenvalues of -1 or 1 implies the function is odd or even respectively. The values of $ \chi (\mathbf{r},\mathbf{r}',\tau) $ and $\langle j(\mathbf{r},\tau) j(\mathbf{r}',0) \rangle $, for the different parity options are shown in Table \ref{par}.

Apart from the four parity options for the operators $P$ and $T$, there can be another kind of behavior. For example, if $\Delta_+(\tau)$ is even in time and $\Delta_-(\tau)$ is odd. Such a situation, where the parity in one variable depends on the sign of the other variable suggest that the time and momentum parities are entangled (not necessarily in the quantum mechanics sense). Thus the parity of each individual operator is not defined and we refer to this state as entangled. There could be several possibilities for such a state and the values of $ \chi (\mathbf{r},\mathbf{r}',\tau) $ and $\langle j(\mathbf{r},\tau) j(\mathbf{r}',0) \rangle $, for one of those possibilities are shown in the last line of Table \ref{par}.

The typical case of CDW require that $\Delta_+(\tau)= \Delta_-(\tau)$, i.e. $P$ even, and independent of the sign of the momentum. One can allow to have explicit non-trivial dependence on the sign of the momentum i.e. $\Delta_+(\tau)= - \Delta_-(\tau)$, which means we are considering p-wave CDW. Hence, we have both possibilities and Table \ref{par} represent a general symmetry classification.

\section{Definition of an order parameter for odd time CDW}
If indeed this state represent a new phase, which possess some order, it is described by an order parameter. Since the order parameter, as it is most commonly defined, vanish in the case of odd time correlations, we define a new order parameter. This can be done, for example, by considering the time derivative of the correlation at equal time, i.e. $\tau =0$. Let us define our order parameter as \cite{Abrahams,Dahal}
\begin{equation} \label{opNew}
D_\mathbf{q}(\mathbf{k}) = {d \over d\tau} \Delta_\mathbf{q}(\mathbf{k},\tau)|_{ \tau =0}.
\end{equation}
While $ D_\mathbf{q} (\mathbf{k}) $ does not depend on time it capture some of the dynamics of the state. For any model Hamiltonian $H$, we can use the Heisenberg equation ${d \over d\tau} O(t) = [H,O]$ to write this order parameter as
\begin{equation} \label{opNew2}
D_\mathbf{q} (\mathbf{k}) = \left\langle [H,a^\dagger_{\mathbf{k}+\mathbf{q}}] a_\mathbf{k} \right\rangle.
\end{equation}
In case the part that is most significant is the correlations at close to equal time $ \tau \ll 1 $, it is plausible to approximate this correlations to first order as
\begin{equation} \label{firstordr}
\Delta_\mathbf{q}(\mathbf{k},\tau) \simeq D_\mathbf{q} (\mathbf{k}) \tau
\end{equation}
and then we can learn from the order parameter (\ref{opNew2}) relevant correlations that are responsible for odd time order (\ref{kOPt}).

It can be very instructive to calculate (\ref{opNew2}) for a specific model and a good candidate would be the textbook model for CDW. However, our discussion of CDW, which involved a phenomenological approach and was focused on electron correlation, is somewhat different from the traditional treatment of CDW, which typically includes a microscopic model and is focused on average of phonon operators. Thus we briefly review the traditional formalism and show how it connects to the one we used. 

The traditional derivation of CDW \cite{Gruner1,Gruner2} involves the electron-phonon interaction \begin{equation} \label{hi}
H_I = \sum_{\mathbf{k},\mathbf{q}} g_{\mathbf{q},\mathbf{k}} \left(b^\dagger_{-\mathbf{q}} + b_\mathbf{q} \right) a^\dagger_{\mathbf{k}+\mathbf{q}}a_\mathbf{k}
\end{equation}
where $ b^{(\dagger)}_{\mathbf{q}}$ is creating (annihilating) a boson with momentum $\mathbf{q}$ and $g_{\mathbf{q},\mathbf{k}}$ is a coupling function. Using mean field methods, a self consistent equation can be obtained for $\langle b_\mathbf{q} \rangle$ with a solution having $\langle b_\mathbf{q} \rangle \ne 0$, which represent a periodic distortion in the underlying lattice. The average $\langle b_\mathbf{q} \rangle$ is related to the gap that is opened in the electron spectrum and it is interpreted as an order parameter. From a phenomenological point of view, if one is interested in the electronic structure, it is not necessary to consider the interaction (\ref{hi}), or even to involve phonons, in order to derive an order parameter. One can simply consider periodic spatial modulation $n_\mathbf{q}$ in the density of electrons, as given by (\ref{ncdw}), and write an expression for $n_\mathbf{q}$ in term of averages on creation operators
\begin{equation}\label{nq}
n_\mathbf{q} = e^{i \phi } \int d\mathbf{k} \langle a^\dagger_{\mathbf{k}+\mathbf{q}} a_\mathbf{k} \rangle =  \int d\mathbf{k} \Delta_\mathbf{q}(\mathbf{k}).
\end{equation}
The presence of modulations $n_\mathbf{q} \ne 0$, implies a non vanishing order parameter regardless of the microscopic model but from (\ref{hi}) it is clear that an $\langle b_\mathbf{q} \rangle \sim \Delta_\mathbf{q}(\mathbf{k})$. So the traditional formalism can be regarded as a possible recipe to construct a microscopic model which can explain our phenomenological quantities.   

We can use (\ref{hi}) for getting an explicit expression for $D_\mathbf{q}(\mathbf{k})$. We have
\begin{equation} \label{opNew3}
D_\mathbf{q} (\mathbf{k}) = \left\langle \sum_{\mathbf{q}'} g_{\mathbf{q}',\mathbf{k}+ \mathbf{q}} \left(b^\dagger_{-\mathbf{q}'} + b_{\mathbf{q}'} \right) a^\dagger_{\mathbf{k}+\mathbf{q}+\mathbf{q}'} a_\mathbf{k} \right\rangle.
\end{equation}
In the typical case, which we discussed above, $\mathbf{q}=\mathbf{q}'=2k_f$. Thus the order parameter is given by electron-hole correlation with a momentum difference of $4k_f$. This observation is consistent with the behavior of $ \chi (\mathbf{r},\mathbf{r}',\tau) $ and $\langle j(\mathbf{r},\tau) j(\mathbf{r}',0) \rangle $ calculated above. Indeed this suggests that the hidden order we are interested in is manifested in higher order correlation between electron and lattice degrees of freedom.

\section{Conclusion}

In conclusion, we present the odd time (frequency) CDW and SDW as an extension to the notion of density waves considered to date. Main feature of these odd time density waves is that there are no equal time modulation in the expectation values of spin and charge densities. Yet as is clear from the overall structure, there are nontrivial dynamic correlations. These nontrivial correlations, Eq. (\ref{xi}) and (\ref{jj2}) could serve as a test for the odd time density wave. As such this work represents the extension of odd-frequency state classification to density waves, beyond superconductors and BEC.

It remains to be seen what are the realistic fermion interactions that would enable the odd time density correlations. We also point out that these dynamic correlations with no equal time expectation value do correspond in a way to a ``hidden order" that cannot be observed as the traditional charge/spin modulation. As such it can be a candidate for materials with hidden orders where one observes thermodynamic features of a transition while no order parameter consistent with the transition can be inferred from the current measurements.

{\em Acknowledgements} We acknowledge useful discussions with Sergey Pershoguba and Jonathan Edge. This work was supported by ERC
DM-321031 and US DOE BES E304.

\appendix*

\section{Current-current correlation} \label{japp}

The current operator is given by
\begin{align}
j_{i}(\mathbf{r})&=\frac{1}{2mi}\left(\Psi^{\dagger}(\mathbf{r})\nabla_i\Psi(\mathbf{r})-\Psi(\mathbf{r})\nabla_i\Psi^{\dagger}(\mathbf{r})\right) \nonumber \\
&= \int d\mathbf{k} d\mathbf{k}' e^{i \mathbf{r} (\mathbf{k}' - \mathbf{k})} a^\dagger_{\mathbf{k}} a_{\mathbf{k}'} \frac{\mathbf{k}' + \mathbf{k}}{2m}.
\end{align}
The current-current correlation is given by
\begin{align}
&\langle T_\tau j_i(\mathbf{r},\tau) j_j(\mathbf{r}',0) \rangle = \nonumber\\
&\int \mathcal{K} e^{i \mathbf{r} (\mathbf{k}_1 - \mathbf{k}_2)} e^{i \mathbf{r}' (\mathbf{k}_3 - \mathbf{k}_4)} \frac{\left(\mathbf{k}_1 + \mathbf{k}_2\right)_i}{2m} \frac{ \left(\mathbf{k}_3 + \mathbf{k}_4\right)_j }{2m} \times \nonumber \\
&\langle T_\tau a^\dagger_{\mathbf{k}_1}(\tau) a_{\mathbf{k}_2} (\tau) a^\dagger_{\mathbf{k}_3}(0) a_{\mathbf{k}_4} (0) \rangle.
\end{align}
where  $\int d \mathcal{K} = d\mathbf{k}_1 d\mathbf{k}_2 d\mathbf{k}_3 d\mathbf{k}_4$ means integration on all momentum variables.

Using Wick’s theorem we can write the average on four operators as the product of two averages on two operators. Since we are not considering the situation of Cooper pairs, averages containing two annihilation operators or two creation operators will vanish.

Thus we are left with two ways of pairing the operators: $ \langle T_\tau a^\dagger_{\mathbf{k}_1}(\tau) a_{\mathbf{k}_2} (\tau) \rangle \langle T_\tau a^\dagger_{\mathbf{k}_3}(0) a_{\mathbf{k}_4} (0) \rangle $ and $\langle T_\tau a^\dagger_{\mathbf{k}_1}(\tau) a_{\mathbf{k}_4} (0) \rangle \langle T_\tau a^\dagger_{\mathbf{k}_3}(0) a_{\mathbf{k}_2} (\tau) \rangle$.

The first way of pairing will simply give us $\langle j(\mathbf{r},\tau) \rangle \langle j(\mathbf{r}',0) \rangle $, and we can assume for simplicity that it vanish. The second way of pairing is related to $\Delta_\mathbf{q}(\mathbf{k},\tau)$. We can make this relation concrete with the mapping
\begin{eqnarray}
\mathbf{k} &=& \mathbf{k}_4, \nonumber\\
\mathbf{q} &=& \mathbf{k}_1 - \mathbf{k}_4, \nonumber\\
\mathbf{k}' &=& \mathbf{k}_2, \nonumber\\
\mathbf{q}' &=& \mathbf{k}_3 - \mathbf{k}_2. \nonumber
\end{eqnarray}
Rewriting the current-current correlation above using these variables yields eq (\ref{jj1}) in the main text.

\end{document}